# Anisotropic superconductivity induced at a hybrid superconducting-semiconducting interface

Anand Kamlapure[1], Manuel Simonato[1], Emil Sierda[1], Manuel Steinbrecher[1], Umut Kamber[1], Elze J. Knol[1], Peter Krogstrup[2,3], Mikhail I. Katsnelson[1], Malte Rösner[1*], Alexander Ako Khajetoorians[1,*]

[1] Institute for Molecules and Materials, Radboud University, 6525 AJ Nijmegen, the Netherlands

[2] Center for Quantum Devices, Niels Bohr Institute, University of Copenhagen, 2100 Copenhagen, Denmark

[3] Microsoft Quantum Materials Lab Copenhagen, 2800 Lyngby, Denmark

*corresponding authors: a.khajetoorians@science.ru.nl, m.roesner@science.ru.nl

**Epitaxial semiconductor-superconductor heterostructures are promising as a platform for gate-tunable superconducting electronics. Thus far, the superconducting properties in such hybrid systems have been predicted based on simplified hybridization models which neglect the electronic structure that can arise at the interface. Here, we demonstrate that the hybrid electronic structure derived at the interface between semiconducting black phosphorus and atomically thin films of lead can drastically modify the superconducting properties of the thin metallic film. Using ultra-low temperature scanning tunneling microscopy and spectroscopy, we ascertain the moiré structure driven by the interface, and observe a strongly anisotropic renormalization of the superconducting gap and vortex structure of the lead film. Based on density functional theory, we attribute the renormalization of the superconductivity to weak hybridization at the interface where the anisotropic characteristics of the semiconductor band structure is imprinted on the Fermi surface of the superconductor. Based on a hybrid two-band model, we link this hybridization-driven renormalization to a weighting of the superconducting order parameter that quantitatively reproduces the measured spectra. These results illustrate the effect of interfacial hybridization at superconductor-semiconductor heterostructures, and pathways for engineering quantum technologies based on gate-tunable superconducting electronics.**

Hybrid superconductor-semiconductor heterostructures are a pathway toward tailoring superconductivity as well as for creating a platform for topological quantum computing (*1, 2*). While the changes in electronic structure created by coupling the bulk bands of a superconductor and a semiconductor can modify both the electronic properties in the semiconductor (*3, 4*), as well as the properties of the superconductor, the interface itself can also introduce a new electronic degree of freedom that is often neglected. Toward this end, it has been shown that superconductivity can be enhanced at the interface between dissimilar materials (*5-7*). For example, the monolayer of iron selenide grown on strontium titanate, a large bandgap semiconductor, exhibits high temperature superconductivity (*5*), with a critical transition temperature nearly an order of magnitude larger than in its parent bulk compound (*8*). Likewise, the interface at lanthanum aluminate on strontium titanate exhibits superconductivity in a 2D electron gas residing at the interface between the two insulating compounds (*6*). While these examples clearly show that superconductivity can be strongly enhanced by an interface, the mechanism for superconductivity in these heterostructures is still strongly debated making it extremely challenging to understand the essential phenomena responsible for these changes (*7, 9*).

Superconductor-semiconductor heterostructures, derived from elemental superconductors, instead provide a tunable platform toward quantifying the various influences of the heterostructure on superconductivity (*10, 11*). So far, the emphasis of numerous studies has been in establishing the interplay between dimensionality and superconductivity, when compared to their bulk counterparts (*12-16*). For example, it has been shown that quantum confinement can modulate $T_c$, through quantum well states, leading to the demonstration of 2D superconductivity (*12-14*). These approaches rely on supporting semiconductors with a sizeable electronic band gap, which largely suppress unwanted hybridization from the semiconductor that otherwise may lead to quasiparticle poisoning. With the growing interest in quantum technologies based on superconductor-semiconductor heterostructures (*1, 17*), it remains unknown how hybrid band structures created at the junction of two materials can be used to controllably tune superconductivity in lower dimensional structures.

Here, we demonstrate the formation of a hybrid superconductor formed between the interface of a superconducting ultra-thin film of lead and semiconducting black phosphorus. The hybrid band structure induced by the interface renormalizes the superconductivity of the lead layers yielding strongly anisotropic characteristics. Using ultra-low temperature scanning tunneling microscopy (STM) and spectroscopy (STS) down to mK temperatures, we quantify the superconducting gap, which shows an unexpected anisotropic structure, as a function of thickness of the lead film. Using spatial imaging in magnetic field, we quantify the resultant Abrikosov lattice which is derived from a strongly anisotropic vortex structure. We observe concomitantly the presence of a strong moiré lattice, which allows us to ascertain the structure of the film, including the induced strain driven by the interface. Using this determined structure as input, we performed *ab initio* calculations based on density functional theory on a minimal moiré structure and observe that the black phosphorus bands influence the Fermi surface of lead in particular regions in *k*-space. Namely, the presence of the interface leads to modifications in the quantum confined states of the lead, in accordance with the symmetry of the interface which we compare to spectroscopic measurements. We developed a hybrid two-band superconducting model which illustrates that superconductivity can be sculpted across regions of the Fermi surface by considering weak hybridization of the lead bands, including the quantum well sub-bands, with the anisotropic black phosphorus bands. This enabled us to fit and discern the detailed gap structure measured, and to extract the *k*-dependent weighting functions of the superconducting quasiparticles. Based on these results, we observed a gradual weakening of the interfacial effect in thicker films, allowing us to distinguish between an ultra-thin limit where the hybrid superconductivity is strongly driven by the interfacial band structure from a thick film limit where the interface presents a weaker perturbation.

Black phosphorus (BP) is a narrow band gap semiconductor ($E_G$ = 0.3 eV) that exhibits a strongly anisotropic electronic dispersion (*18-21*). BP cleaves with an atomically flat surface over macroscopic length scales, exposing the (001) surface. Closed Pb(111) films grow epitaxially on the clean surface of cleaved BP when utilizing a two-stage growth process (see supporting text S1 for two-stage growth) as depicted in Fig. 1A. Using constant-current STM imaging (Fig. 1B), we observed that the absolute thickness in monolayers (*N*) of a given film can be identified using holes in the film which penetrate directly to the BP surface (fig. S1 and fig. S3A). This can be combined with the unique layer-dependent spectroscopic fingerprinting of the local density of states (LDOS), which is discussed later. The typical Pb film exhibits monolayer variations in absolute height, with regions from 7 to 11 monolayers (ML) illustrated (fig. S1). This is in stark contrast to the expected bilayer height variations driven by the quantum well states common to the growth of Pb films on numerous other surfaces (*22-27*). The quenching of the quantum size effect in these Pb films provides a first indication of a hybrid electronic structure that is different than the well-studied Pb films grown on Si.

High-resolution constant-current imaging on various film thicknesses reveals a persistent and strong moiré lattice (Fig. 1C) in addition to atomic resolution of the Pb(111) film with the expected symmetry. The moiré lattice exhibits a rectangular unit cell (5.03 nm x 7.35 nm) which extends along the zig-zag direction of the BP [010] rows. We observed the same moiré pattern, regardless of sample and film thickness, from 3 to 30 ML. We simulated the moiré lattice considering both the interference between the Pb and BP atomic lattices, as well as the strain induced in the Pb film (see supporting text S2 and fig. S2). Based on this, we reproduce the experimentally observed images (Fig. 1D), which confirms the atomic structure of the Pb film relative to the underlying BP. Based on these simulations, we find that all observed films exhibit a uniaxial strain (compressed) of ~ 1% along the [110] direction of the Pb lattice. As we do not observe a wetting layer and all the films show islands and vacancy islands, the strain is likely relieved in [111] direction of the Pb.

We quantify the superconducting properties of these Pb films as a function of thickness, using STS down to $T$ = 30 mK (Fig. 2A). Spectra measured at various locations on a given terrace reveal a uniform superconducting gap of roughly $\Delta$ ~ 1.29 meV. This varies in comparison to the bulk gap value and we do not observe the double gap structure seen on the surface of bulk Pb (*28, 29*). $\Delta(N)$ for 7-9 ML is shown in Fig. 2A and can be characterized by a strongly broadened and V-shaped gap structure, beyond a full superconducting gap. This is different to the superconducting gap of Pb films grown on Si(111), which we measured at similar conditions. The overall modifications to the gap structure driven by the interface, can be described by three signatures each with different relative contributions: (i) a broadening of the conductance near the coherence peaks, (ii) a suppression of the coherence peaks and (iii) a gradual and rounded onset of conductance at the gap edge. These various observations cannot be explained by temperature broadening of a BCS gap, which will evenly broaden the total gap structure. Also, the gap cannot be properly fitted with conventional formulas based on BCS theory (*30, 31*). Therefore, the most natural consideration would be to account for an induced anisotropic gap due to the BP interface, namely an angle dependent superconducting gap, resulting from a renormalization of the Fermi surface from circular to ellipsoidal (*32, 33*). At sufficiently low temperature, an ellipsoidal Fermi surface should lead to a nontrivial modification to the gap structure (e.g. kinks) (*33*). This was not observed in the temperature-dependent experimental spectra and thus motivates the consideration of *k*-dependent weighting of the gap function, as we discuss later. Unlike the expected bilayer oscillation in Δ(*N*) seen for Pb/Si (*12-14*), the overall gap structure and Δ(*N*) do not show any pronounced changes or layer-dependent oscillations and remain robust for a given film. While the variations in the gap are small for a given sample, we observed strong variations between synthesized films, which we discuss later. Likewise, we did not observe strong spatial variations in the superconducting gap depending on the probed location with respect to the underlying moiré lattice. We also did not observe any proximity induced gap in the BP, which could be directly measured through holes in a given Pb film (fig. S3). Finally, we observe a relative difference in the gap structure for ultra-thin films (< ≈12 ML), compared to films approximately 30 ML thick. For the latter, we observed gap structure that still exhibit a degree of anisotropy, but more closely resemble the gap structure of Pb/Si(111). This suggests the interface electronic structure is responsible for the observed variations.

In order to confirm the anisotropic character of the superconducting order parameter, we studied spatially resolved maps of the resulting vortices in an applied out of plane magnetic field for as grown 7.3 ML Pb film. For $B_\perp = 50$ mT, we observed a pronounced triangular Abrikosov lattice (Fig. 2B) with an average vortex-vortex distance of $d_v \sim 217$ nm. High resolution imaging reveals that this vortex lattice is composed of strongly anisotropic vortices (Fig. 2C), which strongly deviate from circular symmetric vortices seen in thin Pb films (*34-36*). Likewise, the vortex structure does not resemble a simplistic ellipsoidal structure seen in other superconductors described by an anisotropic Fermi surface (*32, 33*). These strongly anisotropic vortices vary slightly in their shape and size, most likely due to weak disorder resulting from variations in the film thickness. They exhibit clear deviation from the isotropic vortex structure seen in type-II Pb films (*34-36*), which can be best described as showing reduced symmetry from the hybrid interface. This is in stark contrast to isotropic vortices previously observed for Pb films of the same thickness, synthesized on larger band gap semiconductors (*34, 35*). As shown in Fig. S4, we observed a transition toward nearly isotropic vortex structures in thicker films (~30 ML), which agrees with the evolution toward a more isotropic gap structure for thicker films, seen in Fig. 2. These observations indicate that, in ultra-thin films, the order parameter is strongly influenced by the hybrid structure formed from the BP interface.

By performing spatially dependent spectroscopy along a given vortex, the anisotropic character can be quantified (Fig. 2D). The d*I*/d*V* spectra measured along one of the extended stripes shows a small splitting of the zero bias peak away from the center and extends over several 10's of nanometers (direction 2) before abruptly merging into coherence peaks. However, d*I*/d*V* spectra measured in a nearly perpendicular direction (direction 1) show a splitting of the zero bias peak and smooth merging into the coherence peaks above 10 nm. We note that vortices were observed on all films, while the formation of a clear Abrikosov lattice was only seen on thin films with a sufficiently low defect density. A comparison between a zero bias conductance map and the topography shows no strong correlation. However, a slight distortion of the Abrikosov lattice (especially at B⊥ = 80 mT) indicates weak pinning of the vortices (Fig. S5). Based on magnetic field dependent spectroscopy measurements, we extract $H_{C2} \approx 280$ mT and a coherence length of $\xi \sim 30$ nm (Fig. S5).

In order to uncover the role of the electronic structure on the observed anisotropic superconductivity, we performed density functional theory (DFT) calculations of both a quasi-free standing Pb film as well as the identical film coupled to the BP substrate (See supporting text S3). The mismatch between the BP and Pb lattice constants results in the formation of a moiré pattern. To proceed with the band structure calculations, we thus need to find a reasonably good commensurate approximation with not too large periodicity. To this end, we utilized a minimal moiré model laterally consisting of two √3 Pb unit cells at the experimentally determined 1% strain (Fig. 1C) on three BP unit cells with three BP layers thickness. Periodic boundary conditions force us to strain the BP substrate unit cells. The positions of the Pb atoms are subsequently relaxed. In Fig. 3A-B we present the resulting Fermi surfaces of a 5 ML Pb film unfolded to the primitive Pb Brillouin zone. Next to the quantum well states around Γ we observe a prominent inner hexagonal Fermi surface. These states result from the same Pb band of mixed $p_z$/ $p_x$/ $p_y$ character, which can efficiently hybridize with the anisotropic BP states. This hybridization imprints clearly visible modulations to the hexagonal Fermi surface characterized by a slight and strong inward bending in the $k_y$ and $k_x$ directions, respectively. This underlines the two-fold symmetry imprinting to the Pb states from the BP substrate.

To compare the model DFT calculations with experiments, we measured the local density of states (d*I*/d*V*) as a function of applied voltage (Fig. 3C). Corresponding DFT calculations for 3 to 7 ML films are shown in Fig. 3D. (see fig S8 for more experimental curves and detailed explanation of the DFT spectra). The experimental LDOS exhibits identifiable features reminiscent of the bilayer oscillation of the quantum well states (QWS) for Pb films, which are heavily broadened in comparison to the QWS observed for Pb films on larger gap semiconductors (*13, 37*). In a simple quantum well picture, weak hybridization of the quantum well boundary with the underlying BP leads to delocalization of the confined wavefunction and weakens the confinement effect yielding additional density of states contribution from the hybrid band structure. There are additional features in the LDOS beyond the typical peaks associated with the onset of the subbands. All these features compare well to the calculated DOS (Fig. 3D), where hybridized QWS as well as additional electronic structure introduced by the hybridized heterostructure can be identified. By increasing the film thickness to 28 ML, we reach a regime where the QWS show more expected bilayer oscillation behavior (fig. S8) with some remaining influence from the interface in form of a weak broadening. This observation together with the aforementioned observations regarding the superconductivity in thicker films confirms that by increasing the film thickness, the Pb films still host quantum well states, which are, however, only weakly perturbed by the BP interface as compared to the ultra-thin limit.

The Bogoliubov-de-Gennes (BdG) quasiparticles will be dramatically affected by the hybridization of the two subsystems, especially around the Fermi surface in the regions of the Brillouin zone where the Pb electronic structure is mostly affected by the BP substrate. Based on this, we developed a hybrid two-band superconductor model which considers an isotropic superconducting band (Pb), which can interact via hybridization with a highly anisotropic non-superconducting band (BP) (see supporting text S5). To this end, we allowed only pure Pb states to form Cooper pairs, while the BP states can only hybridize with the Pb ones on a single-particle level. In such a model, we can consider the *k*-dependent effects of hybridization to the Fermi surface and its impact on the BdG quasiparticle spectrum as depicted in Fig. 4A. The main effect related to the experimental data is the emergence of

(i) an anisotropic gap which leads to a stronger pairing potential on certain parts of the Fermi surface, and (ii) an anisotropic density of states which can be described by a so-called weighting function, $w(\theta)$.

From this model, we derived an analytical expression for the superconducting gap which can account for all the relevant details seen in the experimental spectra. The relevant fitting procedure and parameters are discussed in the supporting text S6 and Table S1. In Fig. 4B, we show an experimental spectrum taken on a 7 ML Pb film (also shown in Fig. 2A) and compare it to fits using (a) a conventional isotropic BCS superconductor (magenta) gap function, (b) an anisotropic gap with no weighting function (orange) and (c) an anisotropic gap with a weighting function derived from the proposed hybrid two-band model (red). We find that the two-band model can account for the three measured effects described in Fig. 2, enabling quantification of both $\Delta(\theta)$ and $w(\theta)$, and their influence on the gap shape (Fig. 4C). By considering anisotropic effects alone, at low enough temperature, a step is expected to emerge in the spectral function which is not observed experimentally (see inset of Fig. 4B). Due to $w(\theta)$ the gap opening in the BdG quasiparticle dispersion is strongly suppressed in regions comparable to the hybridization points seen in the DFT calculations from Fig. 3B. Such weighting functions have been used in previous studies to phenomenologically describe observations of high temperature superconductors (*38, 39*). Here, we can directly derive the weighting behavior from the weak hybridization of states at the interface.

We fitted all measured experimental data obtained at $T$ = 30 mK (Fig. 4D) and $T$ = 1.3 K (fig. S9 and fig. S10) with this method. We extracted $\Delta(N)$ for each synthesized film and found that $\Delta(N)$ is nearly thickness independent and does not show a clear bilayer oscillation. Furthermore, we found similar values for the fitting parameters for both temperatures (Table S1). This is not the case for anisotropic fitting models that neglect the weighting function. Additionally, we observed a clear variation in $\Delta$ from sample to sample. This might be explained by small variations in the pristine *p*-doping of the cleaved BP samples, which showed variations in the band edges of pristine BP samples of up to 40 meV for the valence band edge and 20 mV for the conduction band edge (see fig. S11).

In conclusion, we demonstrated the creation of hybrid superconductivity driven by the interfacial coupling between Pb quantum films and BP. The signature of hybrid superconductivity is manifested by the anisotropic renormalization of the superconducting gap and its vortex structure. The anisotropic character of the superconducting properties can be traced to the anisotropic electronic structure of BP, which weakly imprints itself on the hybrid Fermi surface and delocalizes the confined Pb wavefunctions. This weak hybridization results in a weighted distribution of the superconducting quasiparticles, where the order parameter develops a $k$-dependent amplitude. We quantify this weighting through a newly developed hybrid two-band model, which enables direct and quantitative fitting to the gap structure and allows for the extraction of the angle-dependent weighting function. It remains to be understood how the detailed moiré structure is responsible for the resultant superconducting behavior, including the complex vortex structure observed in our experiments. We observe that near 30 ML thickness of the Pb film this behavior is suppressed, distinguishing a limit where we have interface-driven superconductivity, from a limit where the interface is a weaker perturbation to the band structure of lead. The hybrid behavior observed here demonstrates that the interfacial electronic structure provides a new degree of freedom toward sculpting the electronic properties of hybrid superconductor-semiconductor heterostructures, which goes beyond considering the coupling of bulk bands between dissimilar materials. These conclusions play an important role in the development of quantum technologies, which rely on lower dimensional semiconductor nanostructures weakly coupled to superconducting films.

**Acknowledgements**

The experimental part of this project was supported by the European Research Council (ERC) under the European Union's Horizon 2020 research and innovation programme (grant no. 818399). We also acknowledge support from the NWO-VIDI project 'Manipulating the interplay between superconductivity and chiral magnetism at the single-atom level' with project no. 680-47-534. M.I.K. acknowledges support by European Research Council via Synergy Grant 854843 - FASTCORR. We also acknowledge funding from Microsoft Quantum. For the theoretical part of this work,

**Author contributions**

A.K., E.S, M.St., U.K., and E.J.K. performed the experiments. M.R. and A.A.K. designed the experiments. M.Si., M.I.K., and M.R. performed the *ab initio* calculations and developed the hybrid two-band model. A.K., E.S., E.J.K., and A.A.K. performed the experimental analysis. All authors participated in the scientific discussion of the results, as well as participated in writing the manuscript. M.R. acknowledges helpful discussions with Alexander Rudenko.**References**

1. V. Mourik *et al.*, Signatures of Majorana Fermions in Hybrid Superconductor-Semiconductor Nanowire Devices. *Science* **336**, 1003 (2012).
2. S. Nadj-Perge *et al.*, Observation of Majorana fermions in ferromagnetic atomic chains on a superconductor. *Science* **346**, 602 (2014).
3. A. E. G. Mikkelsen, P. Kotetes, P. Krogstrup, K. Flensberg, Hybridization at Superconductor-Semiconductor Interfaces. *Physical Review X* **8**, 031040 (2018).
4. G. W. Winkler *et al.*, Unified numerical approach to topological semiconductor-superconductor heterostructures. *Physical Review B* **99**, 245408 (2019).
5. Q.-Y. Wang *et al.*, Interface-Induced High-Temperature Superconductivity in Single Unit-Cell FeSe Films on $SrTiO_3$. *Chinese Physics Letters* **29**, 037402 (2012).
6. N. Reyren *et al.*, Superconducting Interfaces Between Insulating Oxides. *Science* **317**, 1196 (2007).
7. Z. Li *et al.*, Atomically Thin Superconductors. *Small* **17**, 1904788 (2021).
8. F.-C. Hsu *et al.*, Superconductivity in the PbO-type structure α-FeSe. *Proceedings of the National Academy of Sciences* **105**, 14262 (2008).
9. D. Huang, J. E. Hoffman, Monolayer FeSe on $SrTiO_3$. *Annual Review of Condensed Matter Physics* **8**, 311-336 (2017).
10. S. M. Frolov, M. J. Manfra, J. D. Sau, Topological superconductivity in hybrid devices. *Nature Physics* **16**, 718-724 (2020).
11. M. Benito, G. Burkard, Hybrid superconductor-semiconductor systems for quantum technology. *Applied Physics Letters* **116**, 190502 (2020).
12. Y. Guo *et al.*, Superconductivity Modulated by Quantum Size Effects. *Science* **306**, 1915 (2004).
13. D. Eom, S. Qin, M. Y. Chou, C. K. Shih, Persistent Superconductivity in Ultrathin Pb Films: A Scanning Tunneling Spectroscopy Study. *Physical Review Letters* **96**, 027005 (2006).
14. S. Qin, J. Kim, Q. Niu, C.-K. Shih, Superconductivity at the Two-Dimensional Limit. *Science* **324**, 1314 (2009).
15. T. Zhang *et al.*, Superconductivity in one-atomic-layer metal films grown on Si(111). *Nature Physics* **6**, 104-108 (2010).
16. D. Qiu *et al.*, Recent Advances in 2D Superconductors. *Advanced Materials* **33**, 2006124 (2021).
17. M. T. Deng *et al.*, Majorana bound state in a coupled quantum-dot hybrid-nanowire system. *Science* **354**, 1557 (2016).
18. J. Kim *et al.*, Observation of tunable band gap and anisotropic Dirac semimetal state in black phosphorus. *Science* **349**, 723 (2015).
19. D. A. Prishchenko, V. G. Mazurenko, M. I. Katsnelson, A. N. Rudenko, Coulomb interactions and screening effects in few-layer black phosphorus: a tight-binding consideration beyond the long-wavelength limit. *2D Materials* **4**, 025064 (2017).
20. B. Kiraly *et al.*, Anisotropic Two-Dimensional Screening at the Surface of Black Phosphorus. *Physical Review Letters* **123**, 216403 (2019).
21. B. Kiraly, N. Hauptmann, A. N. Rudenko, M. I. Katsnelson, A. A. Khajetoorians, Probing Single Vacancies in Black Phosphorus at the Atomic Level. *Nano Letters* **17**, 3607-3612 (2017).
22. Y. Jia, B. Wu, H. H. Weitering, Z. Zhang, Quantum size effects in Pb films from first principles: The role of the substrate. *Physical Review B* **74**, 035433 (2006).10

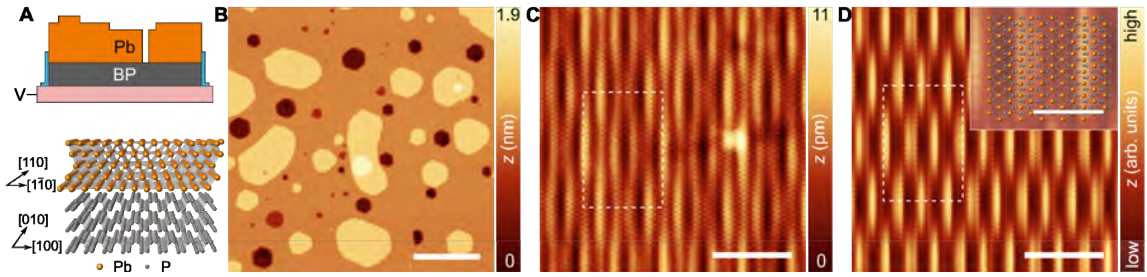

**Fig. 1 Hybrid superconducting heterostructure of lead and black phosphorus.** (**A**) Top: Sketch of the sample geometry showing the cross section of the ultra-thin Pb(111) film and the BP crystal. The blue patches on the two sides represent side contacts. Bottom: Crystal structure of Pb and BP showing the relative orientation between the Pb lattice and the BP lattice. (**B**) Constant-current STM image of a 9 ML Pb film ($V_s$ = 600 mV, $I_t$ = 10 pA, $T$ = 1.3 K, scale bar = 50 nm). (**C**) Atomically resolved constant-current STM image measured on a 7 ML Pb film showing the persistent moiré structure between Pb and BP The white dashed rectangle represents moiré supercell. ($V_s$ = 100 mV, $I_t$ = 100 pA, $T$ = 30 mK, scale bar = 5 nm). (**D**) Simulated moiré structure as described in supporting text S2. The white dashed rectangle represents moiré supercell (scale bar = 5 nm). Inset: lattice structure of the interface overlayed on the zoomed-in image of the simulated moiré structure (inset scale bar = 2 nm).

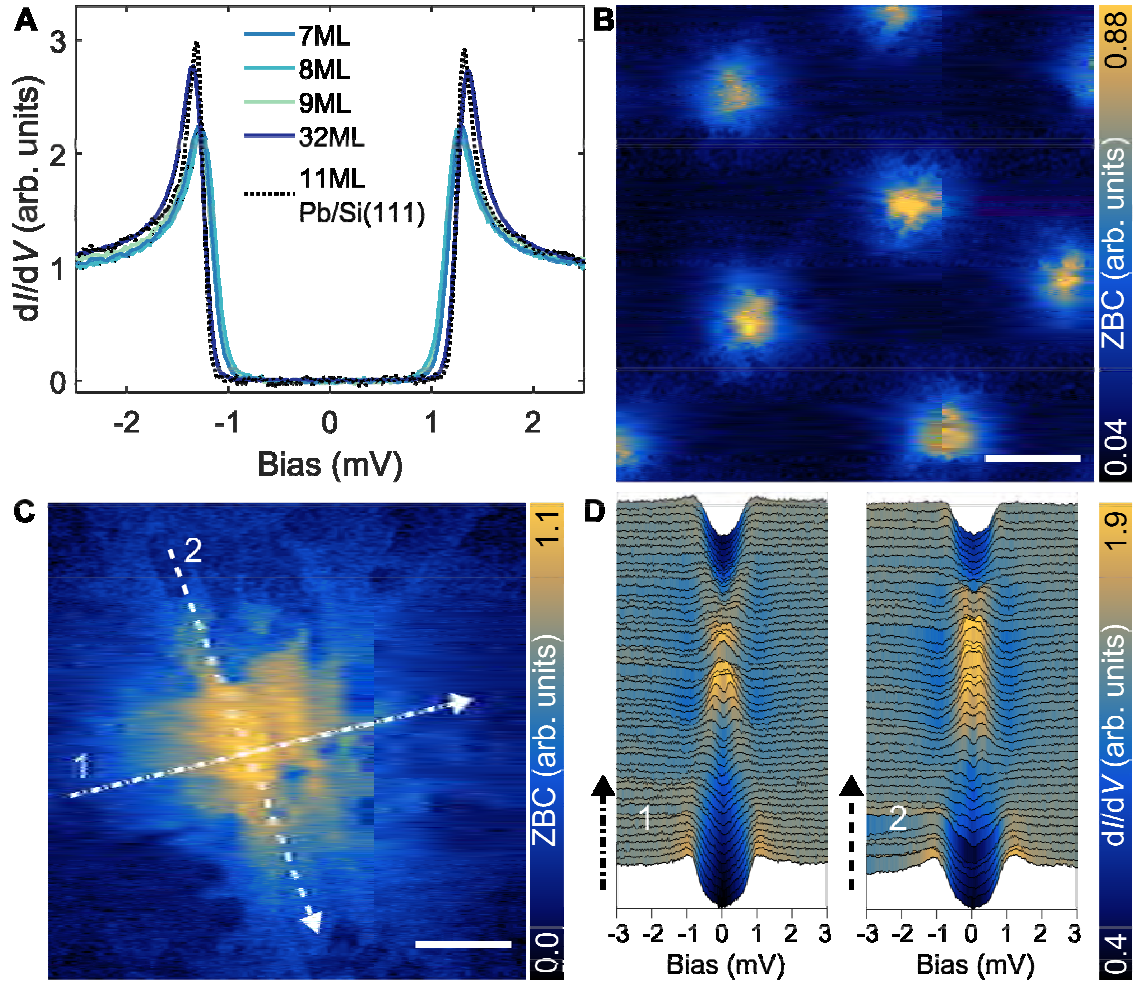

**Fig. 2 Hybrid superconducting gap and anisotropic vortex structure.** (**A**) d$I$/d$V$ spectra measured at $T$ = 30 mK on different layer thicknesses of the Pb film grown on BP, plotted together with the spectrum measured on 11 ML Pb grown on Si(111). For comparison, all the spectra are normalized at $V$ = 3 mV. (Pb/BP: $V_{stab}$ = 5 mV, $I_{stab}$ = 200 pA, $V_{mod}$ = 50 uV; Pb/Si(111): $V_{stab}$ = 5 mV, $I_{stab}$ = 200 pA, $V_{mod}$ = 20 uV). (**B**) Zero bias conductance (ZBC) map measured on as grown 7.3 ML Pb film at $T$ = 30 mK and $H$ = 50 mT showing the Abrikosov vortex lattice (scale bar = 100 nm). Image in (**C**) shows ZBC map of a single vortex (scale bar = 20 nm). Imaging parameters for (B) and (C): $V_{stab}$ = 10 mV, $I_{stab}$ = 10 pA, $V_{mod}$ = 200 uV, $\Delta z$ = - 80 pm. (**D**) d$I$/d$V$ spectra measured across the vortex in two directions (1,2) along the lines through the vortex in (C), each of 84 nm in length. ($V_{stab}$ = 5 mV, $I_{stab}$ = 200 pA, $V_{mod}$ = 50 uV).

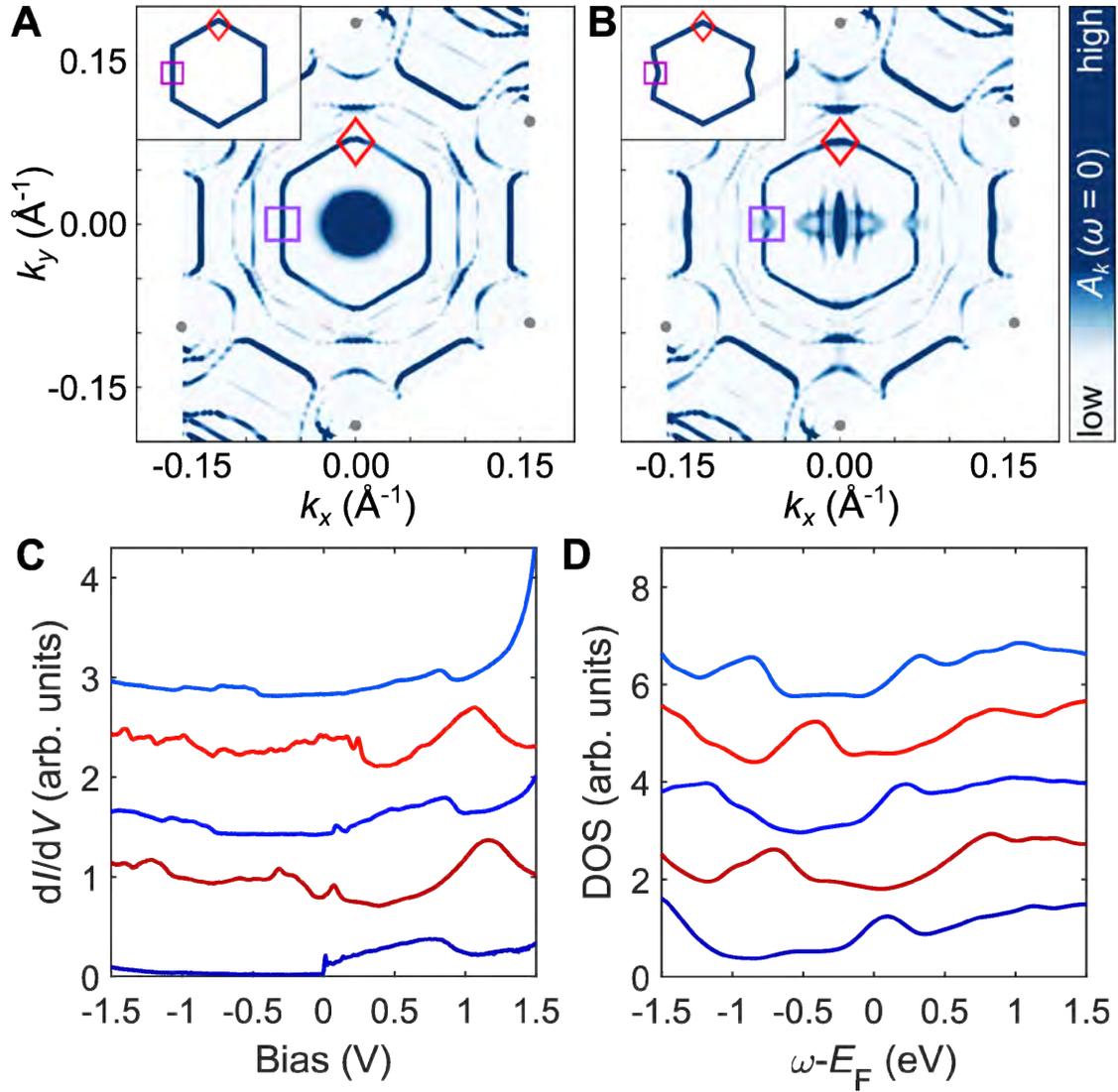

**Fig. 3 Electronic structure calculation of the hybrid structure and comparison with the measured LDOS.** (**A-B**) DFT calculated Fermi surface for 5 ML Pb (**A**) and 5 ML Pb on 3 ML BP (**B**). The insets show a schematic representation of the undistributed inner Fermi surface of 5 ML Pb (**A**) compared to the distorted Fermi surface of 5 ML Pb resulting from the hybridization with underlying BP substrate (**B**). Diamond and square shapes in the inset show the location of K and M points in the corresponding panels. (**C**) d$I$/d$V$ spectra measured in wide bias range on Pb films of various thicknesses from 3 ML (bottom) to 7 ML (top). Films with an even (odd) number of MLs are color coded with red (blue) shades. The spectra are shifted vertically for clarity ($V_{stab}$ = 1.5 V, $I_{stab}$ = 500 pA, $V_{mod}$ = 5 mV, $T$ = 35 mK). Corresponding calculated spectra for Pb films of various thicknesses are plotted in (**D**) with corresponding color shades to the spectra in (**B**).

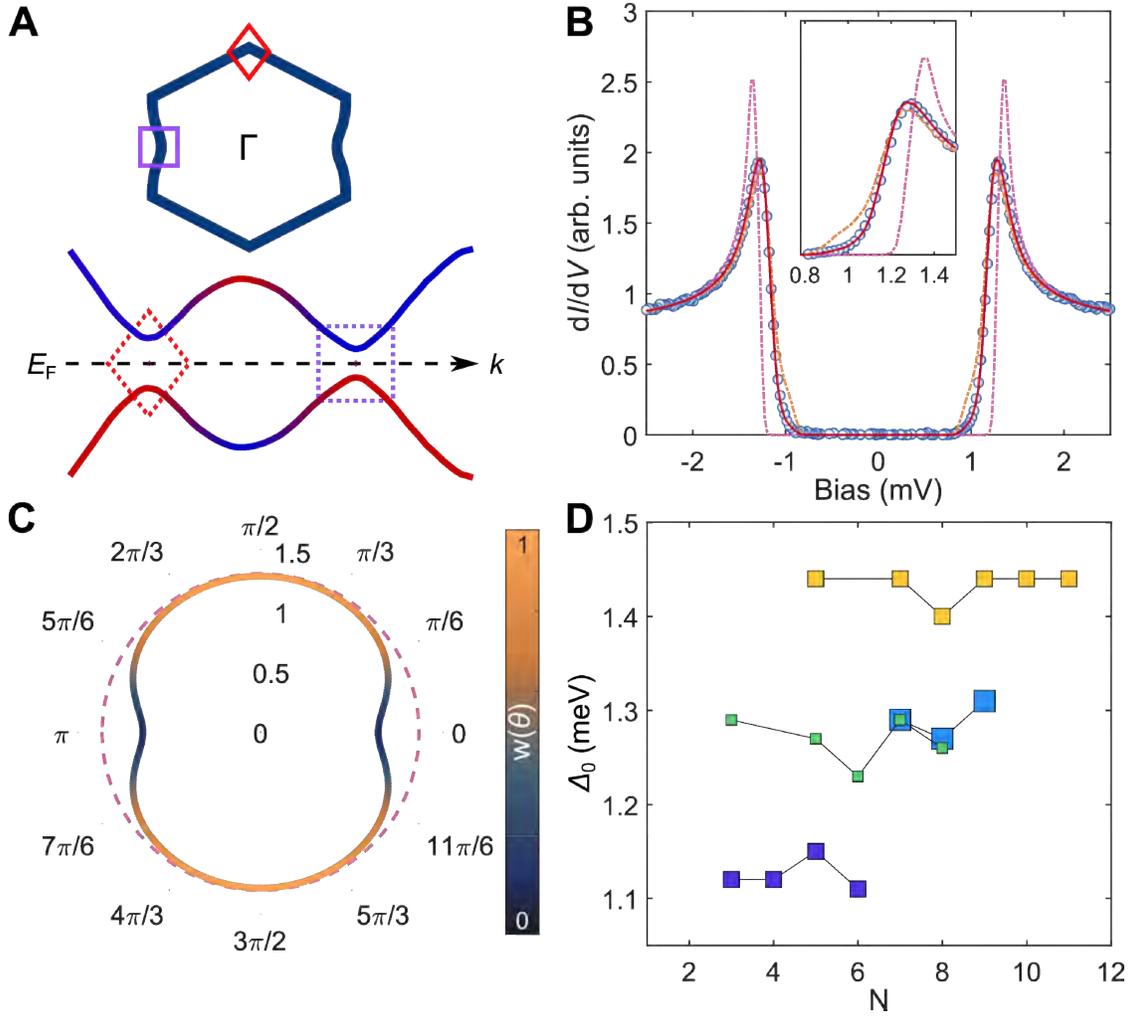

**Fig. 4 Modelling the hybrid superconducting gap and extracting the anisotropic weighting function.** (**A**) Top: sketch of the distorted Fermi surface for Pb/BP showing the K (diamond) and M (square) points. Bottom: Schematic band structure for the hybrid two band model showing the BdG band dispersion and opening of an anisotropic gap. Dashed lines represent hybridized bands in the normal state. Dashed diamond and square shapes in the diagram represent the K and M points of the Fermi surface, respectively. (**B**) d$I$/d$V$ spectrum (open blue circles) measured at $T$ = 30 mK on a 7 ML film (same as in Fig. 2A) together with the fit employing a two-band model (red curve). The fit uses an anisotropic gap and an anisotropic density of states represented by a weight function ($w(\theta)$) as shown in the polar plot in (**C**). The magenta and orange curves in (**B**) represent simulated spectra with an isotropic gap (dashed line in (**C**)) and an anisotropic gap without incorporating the weight function, respectively. (**D**) $\Delta_0(N)$ obtained from fitting plotted for different sample preparations measured at $T$ = 30 mK. The different colors of scatter points represent different preparations.

Supplementary Materials for

# Anisotropic superconductivity induced at a hybrid superconducting-semiconducting interface

Anand Kamlapure[1], Manuel Simonato[1], Emil Sierda[1], Manuel Steinbrecher[1], Umut Kamber[1], Elze J. Knol[1], Peter Krogstrup[2,3], Mikhail I. Katsnelson[1], Malte Rösner[1,*], Alexander Ako Khajetoorians[1,*]

[1] *Institute for Molecules and Materials, Radboud University, 6525 AJ Nijmegen, the Netherlands*

[2] *Center for Quantum Devices, Niels Bohr Institute, University of Copenhagen, 2100 Copenhagen, Denmark*

[3] *Microsoft Quantum Materials Lab Copenhagen, 2800 Lyngby, Denmark*

*corresponding authors: a.khajetoorians@science.ru.nl, m.roesner@science.ru.nl

# Contents

## S1 Materials and methods

All the scanning tunneling microscopy/spectroscopy (STM/S) measurements were performed in two home-built systems with base temperatures of 30 mK (*1*) and 1.3 K (*2*). Both systems are equipped with UHV chambers for sample preparations as well as a cold stage (~100K) for low-temperature growth. In the current study, we use black phosphorus (BP) crystals substrate which cleaves easily along [001] direction as the van der Waals interaction between P layers is weak. Imaging such cleaved surfaces showed atomically flat topography within the scan range. Prior to the Pb film growth, black phosphorous (BP) crystals were first cleaved *in-situ* and transferred to the cold stage for the low temperature deposition. Pb was then evaporated from a Knudsen cell with a constant rate of 0.2ML per minute. After the desired thickness was achieved, the sample was annealed shortly (~ 3-5 min) by resting the sample on the transfer arm at room temperature and subsequently inserted into the STM head for the low temperature measurements. Electronic properties of the samples were then measured using an electrochemically etched W or Cr tip which is coated with Au by dipping into a Au(111) crystal. For the scanning tunneling spectroscopy measurements, a standard lock-in technique was employed where a typical modulation voltage of $V_{\text{mod (rms)}} = 20 - 100$ µV ($f$ = 877 or 893.7 Hz) was added to bias and applied to the sample. To facilitate tunneling at low temperature, we made side contacts on the BP crystals using conducting epoxy. Fig. S1 shows a typical low temperature growth showing closed Pb film and thicknesses from 7ML to 11ML. At a few locations of the sample, the film shows holes all the way to the BP crystal.

## S2 Simulation of the moiré pattern

To simulate the experimentally observed moiré patterns, we first generated a real space image of a single layer of Pb(111) and the topmost layer of a BP(001) lattice (*3, 4*). We used the following function to generate a hexagonal lattice for Pb(111).

$$f_{Pb} = \frac{1}{9} + \frac{8}{9}\left[\cos\left(\frac{1}{2}\boldsymbol{k}_1\boldsymbol{r}\right)\cos\left(\frac{1}{2}\boldsymbol{k}_2\boldsymbol{r}\right)\cos\left(\frac{1}{2}\boldsymbol{k}_3\boldsymbol{r}\right)\right]. \tag{1}$$

Here, $\boldsymbol{k}_n\boldsymbol{r} = xk\cos\left(\theta + \frac{n\pi}{3}\right) + yk\sin\left(\theta + \frac{n\pi}{3}\right)$, where n = 1-3, $\theta$ is the rotation angle for the lattice ($\theta = 0$ in our case) and $k = \frac{2\pi}{(\sqrt{3}/2)a_{Pb}}$, with $a_{Pb}$ = lattice constant for Pb = 3.5 Å.

For the BP lattice we used the following function which generates two offset rectangular sublattices mimicking the top surface of the BP crystal:

$$f_{BP} = p_1 + p_2[\cos(xk_b) + \cos(yk_a)]^m + p_3[\cos(xk_b + d) + \cos(yk_a + a/2)]^n. \tag{2}$$

Here, $k_a = \frac{2\pi}{a}$, $k_b = \frac{2\pi}{b}$, with $a$ = 3.313 Å, $b$ = 4.374 Å, $d$ = sublattice offset = 1.48 Å, $p_1$ = constant offset, $p_{2,3}$ = coefficient for the two offset lattices and $m, n$ = power indices are for the two sublattices that are needed to generate a corrugation that is comparable to experimental STM topograph. The two lattices were generated with a size of 40 nm x 40 nm on a grid with 2000 x 2000 pixels. Fig. S2(A,B) shows the generated Pb(111) and BP(001) lattices.

The resulting moiré pattern was then calculated as a convolution of both lattices,

$$M = f_{Pb} \times f_{BP}. \tag{3}$$

Before comparing resulting pattern with experiments, we smoothed the generated image using a two-dimensional Gaussian filter. The resulting moiré image is shown in Fig. S2C. We could capture all the essential features such as the moiré supercell size, vertically elongated shapes, etc., by adding ~1% of strain in the Pb lattice along the zigzag direction of the BP lattice, using the following parameters: $p_2 = 0.12$, $p_3 = 0.2$, $m = 1$ and $n = 2$.

### S3 *Ab Initio* calculations

All *ab initio* calculations were performed using the Vienna Ab Initio Simulation Package (VASP)(*5, 6*) utilizing the PBE exchange functional (*7*) within Projector Augmented Wave basis sets (PAW)(*8, 9*). The energy cutoff was set to 318.8 eV and we used for all (supercell) calculations 12x12x1 (3x12x1) k grids.

### Supercell geometry

To simulate heterostructures consisting of several monolayers of Pb on the BP substrate we construct minimal lateral supercells with six primitive Pb unit cells per four BP unit cells. This corresponds to three rectangular Pb unit cells (spanned by $a_x \times a_y$) on top of four BP unit cells (spanned $b_x \times b_y$). A top view of such a structure consisting of one monolayer Pb on a BP monolayer is depicted in Fig. S6A. For the construction of these cells, we first optimized the lateral lattice constant of a five-layer Pb film yielding $a_0 \approx 3.57$ Å Afterwards we applied the experimentally verified 1% strain in *y*-direction ([110] direction of the primitive Pb triangular lattice) yielding $a_x \approx 6.204$ Å and $a_y \approx 3.534$ Å. To model the bulk BP substrate, we use three monolayer of BP separated by $b_z = 5.3$ Å and set its lattice constant in *y*-direction to $b_y = a_y$ to match the minimal rectangular Pb unit cell in this direction, as depicted in Fig. S6B. The lattice constant of BP in *x*-direction was set to $4b_x = 3a_x$ to create the minimal moiré cell. This way we get numerically feasible supercells with 78 (90) atoms for five (seven) monolayers of Pb while applying a strain of about 1 - 3.3/3.534 $\approx$ 6.6% and 1- 4.601/4.653 $\approx$ 1.1% in *y* and *x* direction to the BP substrate (*10*). The Pb atomic positions on the fixed BP substrate are finally relaxed until all forces acting on Pb atoms were smaller than 2.5 meV/Å. This yields an average Pb-BP separation in z-direction of $d \approx 3.3$ Å. To minimize spurious interactions in *z*-direction due to the applied periodic boundary conditions we use a supercell height of $h \approx 45$ Å.

### Unfolded spectra

To analyze the Fermi surfaces and band structures of the heterostructures we unfold the spectral functions of the supercells to the Brillouin zone of the primitive triangular unit cell of the Pb films using the approach from Ref. (*11*) according to

$$A(k, \omega) = \sum_m P_{Km}(k)\delta(E_m - \omega) \quad \text{with} \quad P_{Km}(k) = \sum_n |\langle Km|kn\rangle|^2. \tag{4}$$

Here, $|kn\rangle$ and $|Km\rangle$ are Kohn-Sham states of the primitive and super cells, respectively. For the unfolded Fermi surfaces shown in the main text, we plot $A(k, \omega \approx E_f)$ as a colormap for the whole primitive first Brillouin zone. In Fig. S7 we additionally show the resulting band structures, where the dot size corresponds to the value of $A(k, \omega)$ at the given $k$-point and energy $\omega$. To analyze the resulting spectral function in comparison to the experimental STS spectra, we finally define the quasi-local spectral function as

$$\tilde{\rho}(\omega) = \sum_k g(k)A(k, \omega) \quad \text{with} \quad g(k) = \frac{1}{\sigma_k\sqrt{2\pi}} exp\left(-\frac{1}{2}\frac{k^2}{\sigma_k^2}\right) \tag{5}$$

with $\sigma_k$ defining the range of significant $k$ points around Γ.

In Fig. S7 we show the unfolded band structures and quasi-local spectral functions for 5 and 6 Pb monolayers. In red we present the free-standing data and in blue the unfolded spectra including the effects of the BP substrate. In the case of 5 Pb monolayer, we can clearly identify the quantum well state as a rather flat band around $E_F$ at Γ, which leaves a clear footprint in $\tilde{\rho}(\omega)$ (for $\sigma_k = 0.01$). Towards *M* and *K* this state starts to disperse by first increasing and afterwards decreasing in energy, whereby it crosses the Fermi level. Due to the finite strain of 1% in *x*-direction a slight asymmetry is imprinted. Upon including the BP substrate, this former well-defined quantum well state is heavily deformed and so are those states, which cross the Fermi level. In detail, by comparing the free-standing (red) band structure to the unfolded one of the heterostructure (blue), we see that a variety of new BP states appear, e.g. between 0 and 0.5 $eV$ between $M$ and $\Gamma$ or between 0.5 and 1.25 $eV$ around $\Gamma$. In the latter range we see three BP bands, which results from using three BP layer as the substrate. Especially the BP and Pb states around the Fermi level hybridize strongly, which significantly broadens the corresponding feature in $\tilde{\rho}(\omega)$. Furthermore, the additional BP states at slightly higher energies around $\Gamma$ gain significant weight in $\tilde{\rho}(\omega)$. For the 6 Pb layer case, we can observe a similar behavior, but here the quantum well state is shifted downward in energy (to around $-0.75$ $eV$ at $\Gamma$) while another one appeared around +1.0eV at Gamma.

## S4 Superconducting spectral function for anisotropic dispersions

Given the dispersion $E_k$ the spectral function is defined by

$$\rho(\omega) = \frac{1}{4\pi^2} \int d\mathbf{k}\delta(\omega - E_k) = \frac{1}{4\pi^2} \int_\Gamma d\gamma \frac{1}{|\nabla E_k|} \tag{6}$$

with $\Gamma$ being the set $k_\Gamma$ of $k$-points satisfying $\omega = E_k$.

For conventional superconductors with $E(k,\theta) = \sqrt{\zeta^2(k,\theta) + \Delta^2(k,\theta)}$ (in two dimensions and in polar coordinates) the range in which the pairing amplitude $\Delta$ significantly differs from zero is generally limited in $k$-space to a narrow stripe around the Fermi surface of width $|\zeta_k| < \omega_D$ with $\omega_D$ being the Debye frequency. We can thus approximate $\Delta(k,\theta) \simeq \Delta(k_F,\theta) \equiv \Delta(\theta)$.

We proceed with a generic anisotropic quadratic dispersion using the effective masses $m_x^* > m_y^*$ and utilizing polar coordinates

$$\zeta_k = \frac{1}{2m_x^*}k_x^2 + \frac{1}{2m_y^*}k_y^2 - \mu = \frac{1}{2m_x^*}k^2 \underbrace{[1 + \epsilon\sin^2(\theta)]}_{=F(\theta)} - \mu. \tag{7}$$

Here, $\mu$ denotes the chemical potential and $\epsilon = \frac{m_x^*}{m_y^*} - 1$. The function $F(\theta)$ characterizes the anisotropy in the dispersion which modulates the Fermi surface as a function of $\theta$ according to the Fermi wavevector

$$k_F(\theta) = \sqrt{\frac{2m_x^*\mu}{F(\theta)}} = \frac{k_0}{\sqrt{F(\theta)}} \tag{8}$$

with $k_0$ being the Fermi wavevector of an isotropic dispersion with effective mass $m_x^*$. With this, the set $k_\Gamma$ that satisfies $E^2(k_\Gamma,\theta) = \zeta^2(k_\Gamma,\theta) + \Delta^2(\theta) = \omega^2$ is given by

$$k_\Gamma^2(\omega,\theta) = k_F^2(\theta) \pm \frac{2m_x^*}{F(\theta)}\sqrt{\omega^2 - \Delta^2(\theta)} \approx k_F^2(\theta). \tag{9}$$

The latter approximation is justified since we are mostly interested in the energy range of $\omega \approx \Delta(\theta)$. With this we can approximate the denominator in Eq. (6) $\nabla E_k$ as

$$\nabla E(k_F(\theta)) \approx \frac{|\zeta(k_F(\theta))|}{\sqrt{\zeta^2(k_F(\theta)) + \Delta^2(\theta)}}|\nabla\zeta(k_F(\theta))| = \frac{\sqrt{\omega^2 - \Delta^2(\theta)}}{|\omega|}|\nabla\zeta(k_F(\theta))|, \tag{10}$$

where we additionally assumed that the $\Delta(\theta)\nabla\Delta(\theta)$ term is negligibly small. With $|\nabla\zeta(k_F(\theta))| \simeq \frac{1}{m_x^*}F(\theta)k_F(\theta) = \frac{1}{m_x^*}k_0\sqrt{F(\theta)}$ we finally obtain

$$\rho(\omega) = \rho_0 \frac{1}{2\pi} \int_0^{2\pi} d\theta \frac{1}{F(\theta)} \frac{|\omega|}{\sqrt{\omega^2 - \Delta^2(\theta)}}. \tag{11}$$

Here, $\rho_0 = \frac{m_x^*}{2\pi}$ denotes the density of states of the isotropic parabolic dispersion relation in 2D. This expression for the superconducting spectral function is vastly reminiscent of the conventional one known for s-wave BCS superconductors, with the important difference that Eq. (11) is also valid for anisotropic Fermi surfaces, which we here account for by introducing the function $F(\theta)$.

**S5 Hybrid two-band model**
To describe the combined heterostructure of the Pb film and the BP substrate, we utilize a minimal two-band model of the form

$$H = \sum_{k,\sigma} \xi_k c_k^\dagger c_k + \sum_{k,k'} V_{kk'} c_k^\dagger c_{-k}^\dagger c_{-k'} c_{k'} + t \sum_{k,\sigma} [c_k^\dagger b_k + \text{h.c.}] + \sum_{k,\sigma} \eta_k b_k^\dagger b_k. \tag{12}$$

Here, $c_k^\dagger$ ($c_k$) and $b_k^\dagger$ ($b_k$) are Pb and BP electron creation (annihilation) operators, $\xi_k$ and $\eta_k$ represent the isotropic and anisotropic electronic dispersions of Pb and BP, respectively, and $t$ describes a finite local hybridization between the distinct states. Most importantly, we allow only within the Pb states for a finite BCS-like Cooper pairing parameterized by $V_{kk'}$. Upon mean-field decoupling of the latter term and by defining the SC order parameter, i.e. the gap function $\Delta_k = \sum_{k'} V_{kk'} \langle c_{-k'} c_{k'} \rangle$, we get

$$H_{\text{MF}} = \sum_{k,\sigma} \xi_k c_k^\dagger c_k + \sum_k [\Delta_k c_k^\dagger c_{-k}^\dagger + \text{h.c.}] + t \sum_{k,\sigma} [c_k^\dagger b_k + \text{h.c.}] + \sum_{k,\sigma} \eta_k b_k^\dagger b_k. \tag{13}$$

Within the extended Nambu-Gorkov space spanned by the spinor $\varphi_k^\dagger = (c_k^\dagger, c_{-k}, b_k^\dagger, b_{-k})$ we can define the Green's function $G$

$$G^{-1} = [i\omega_n 1 - H_{\text{MF}}] = \begin{bmatrix} i\omega_n - \xi_k & -\Delta_k & -t & 0 \\ -\overline{\Delta}_k & i\omega_n + \xi_k & 0 & t \\ -t & 0 & i\omega_n - \eta_k & 0 \\ 0 & t & 0 & i\omega_n + \eta_k \end{bmatrix}. \tag{14}$$

By inverting $G$ and using the Nambu-Gorkov Dyson equation, we can analytically derive the explicit self-consistent gap equation reading

$$\Delta_k = -\sum_{k'} V_{kk'} \frac{\Delta_{k'}}{\sqrt{(E_{k'}^2 - \eta_{k'}^2)^2 + 4t^2[\Delta_{k'}^2 + (\eta_{k'} + \xi_{k'})^2]}} \left[ \frac{\eta_{k'}^2 - \lambda_-^2}{2\lambda_-} \tanh\left(\frac{\beta\lambda_-}{2}\right) - \frac{\eta_{k'}^2 - \lambda_+^2}{2\lambda_+} \tanh\left(\frac{\beta\lambda_+}{2}\right) \right], \tag{15}$$

with $E_k = \sqrt{\xi_k^2 + \Delta_k^2}$ and the full BdG quasiparticle dispersions

$$\lambda_\pm^2(k) = \frac{1}{2}\left[ 2t^2 + E_k^2 + \eta_k^2 \pm \sqrt{(E_k^2 - \eta_k^2)^2 + 4t^2[\Delta_k^2 + (\eta_k + \xi_k)^2]} \right]. \tag{16}$$

In the limit of vanishing hybridization, i.e. $t = 0$, this reduces to the common BCS gap equation, as the two systems, Pb and BP, are fully disentangled. For finite $t$ however, we observe that both the gap equation and the BdG quasi-particle dispersions are affected. Thus, although only Pb states experience a finite pairing potential the resulting superconducting state is decisively affected by the normal but anisotropic BP states.

We proceed with analyzing the new BdG quasiparticle dispersions. To this end, we first introduce the normal dispersion of the hybridized band structure as resulting from the diagonalization of the single-particle terms of our two-band model

$$h_{\pm}(k) = \frac{\xi_k + \eta_k}{2} \pm \frac{1}{2}\sqrt{(\xi_k - \eta_k)^2 + 4t^2} \,. \tag{17}$$

With this, we can expand $\lambda(k, \theta)$ in polar coordinates and in the weak hybridization limit as[1]

$$\lambda_{\pm}^2(k, \theta) \simeq h_{\pm}^2(k, \theta) + \Delta^2(k, \theta), \tag{18}$$

where

$$\Delta^2(k,\theta) = \Delta_0^2 \left[1 - \frac{\eta(k,\theta)t^2}{(\eta(k,\theta) - \xi(k,\theta))^2(\eta(k,\theta) + \xi(k,\theta))}\right]. \tag{19}$$

We evaluate $\Delta$ at the hybridized Fermi surface $k'_F(\theta)$, defined by the condition $h_-(k_F'(\theta)) = 0$. This yields

$$\Delta(\theta) \simeq \Delta_0 \left[1 - \frac{t^2}{\eta^2(k'_F(\theta))}\right] = \Delta_0 \left[1 - \frac{\xi^2(k'_F(\theta))}{t^2}\right]. \tag{20}$$

The approximation results from expansion in the weak hybridization limit, identifying the small parameter $\frac{t}{\eta} \ll 1$, and neglecting higher orders in $t$. Note that in the $t = 0$ limit we correctly recover the effective gap $\Delta = \Delta_0$. From this we see that even in the presence of a constant pairing amplitude $\Delta_k = \Delta$, the hybridization at the interface can lead to an anisotropic effective energy gap $\Delta(\theta)$. Furthermore, the effective gap $\Delta(\theta)$ is always smaller than or equal to the initial gap $\Delta_0$. The minima of the effective gap are in fact to be found where $\xi(k'_F)$ deviates the most from zero, i.e. where the hybridized Fermi surface deviates from the original Fermi surface of the single band in the Pb thin film. In these regions the superconducting states from the Pb band are mostly mixed with *non-superconducting* states from the band in the semiconductor substrate, thereby reducing the energy gap.

With this, we can map the two-band SC model to a single anisotropic SC band with effective anisotropic energy gap. The hybridization between the two bands is then ultimately responsible for both the imprinting of the anisotropic character to the final single band model - therefore determining the weight function $w(\theta)$ - as well as for the anisotropy in the effective gap $\Delta(\theta)$. How the single anisotropic band and the effective gap are related to the two bands is to be determined by the choice of a specific model.

**Weighting function and fitting parameters**
To describe our two-band system, we used an isotropic quadratic dispersion $\xi_k$ for the Pb band and an anisotropic quadratic dispersion $\eta_k$ for the BP band. In polar coordinates we have

$$\xi(k, \theta) = \frac{1}{2m^*}k^2 - \mu \quad and \quad \eta(k, \theta) = \frac{1}{2m_x^*}k^2 F(\theta) + \delta - \mu, \tag{21}$$

---

[1] The expansion parameter is here $\epsilon_k^2 = \frac{4t^2[\Delta^2 + (\eta + \xi)^2]}{(E^2 - \eta^2)^2}$, and we only require it to be small when evaluated at the original Fermi surface, where by definition $\xi = 0$. This condition is therefore satisfied when the energy gap $\Delta$ and the hybridization $t$ are small compared to the absolute value $\eta(k_F)$. We simply require that the two bands do not cross at the Fermi surface of Pb, so that $\eta(k_F)$ is large enough in modulus to render $\epsilon_{k_F}$ small.

where $m^*$ describes the isotropic effective mass of Pb, $F(\theta) = 1 + \epsilon\sin^2(\theta)$ controls the anisotropy in the BP band, $\epsilon = \frac{m_x^*}{m_y^*} - 1$, and $\delta$ represents the energy offset between the Pb and BP bands. For this model, the Fermi wavevector in the hybridized system and in the weak hybridization limit reads

$$k'_F(\theta) = \left[2m^*\mu\left(1 + \frac{t^2}{\mu^2}\frac{1}{\frac{\delta}{\mu} - 1 + \frac{m^*}{m_x^*}F(\theta)}\right)\right]^{\frac{1}{2}} = k_0\left(1 + \frac{p^2}{\sigma - 1 + \chi F(\theta)}\right)^{\frac{1}{2}}, \quad (22)$$

where we introduced the parameters:

$$p = \frac{t}{\mu}, \quad \sigma = \frac{\delta}{\mu}, \quad \chi = \frac{m^*}{m_x^*}. \quad (23)$$

Here $k_0$ denotes the isotropic Fermi wavevector in Pb before hybridization. With this we can identify the weighting function in the hybridization model

$$w(\theta) = 1 + \frac{p^2}{\sigma - 1 + \chi F(\theta)}. \quad (24)$$

With Eq. (20) and the Fermi vector from Eq.(22) we obtain:

$$\Delta(\theta) = \Delta_0\left[1 - \frac{p^2}{[\sigma - 1 + \chi F(\theta)]^2}\right]. \quad (25)$$

## S6 Modeling STS spectra with hybrid two band model

In order to determine the superconducting gap structure, we numerically fit experimental d*I*/d*V* curves to differential tunneling conductance derived as below.

The tunneling current between the tip and sample is given by,

$$I(U,T) \propto \int_{-\infty}^{\infty} \rho_t(E)\rho_s(E + eU)[f(E + eU, T) - f(E, T)]dE, \quad (26)$$

where, $\rho_s(E)$ is the sample density of states, $\rho_t(E)$ is the tip density of states and $f$ is the Fermi function. We used the Maki formalism to obtain the sample density of states (*12, 13*) and modified it to include the anisotropic superconducting gap ($\Delta(\theta)$) and an anisotropic weighting function ($w(\theta)$) given by Eq. (24) and (25),

$$\rho(\theta, \omega) = w(\theta) \cdot \Re\left(\frac{u}{\sqrt{u^2 - 1}}\right)$$
$$\text{with, } u = \frac{\omega}{\Delta(\theta)} + \zeta\frac{u}{\sqrt{1 - u^2}}. \quad (27)$$

Here, $\zeta$ is the pair breaking term which accounts for the weak suppression of the coherence peaks in our system. The resulting differential conductance is given as (*14*),

$$\frac{dI}{dU}(U,T) \propto \iint_{-\pi,-\pi}^{+\pi,+\pi} \sin(\alpha)I(U + \sqrt{2}V_{mod}\sin(\alpha), \theta, T)d\alpha d\theta \quad (28)$$

where, *V*$_{mod}$ is the bias modulation used for the STS measurements. Here, the additional integration over $\alpha$ accounts for the broadening in d*I*/d*V* caused by the modulation voltage.

**Suporting figures**

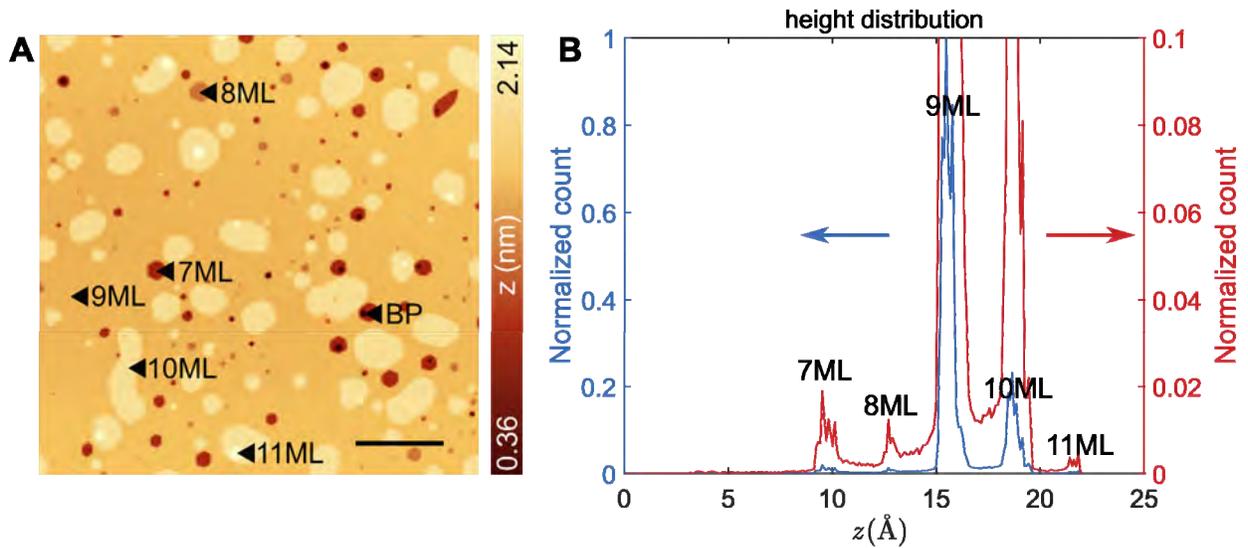

**Fig. S1. Overview of the growth of Pb(111) films on BP(001) and the distribution of film thickness.** (**A**) Constant-current STM image of a typical growth of Pb on BP showing monolayer height variations. The image also shows holes penetrating all the way to BP. ($V_s$ = 600 mV, $I_t$ = 10 pA, $T$ = 1.3 K, scale bar = 80 nm). (**B**) Histogram of the measured heights. Blue and red curves are the same, but plotted on two different y-scales to enhance smaller peaks. Peaks correspond to the different layer thicknesses. From the height distribution, it is clear that the dominant layer thickness for the given growth is 9 ML.

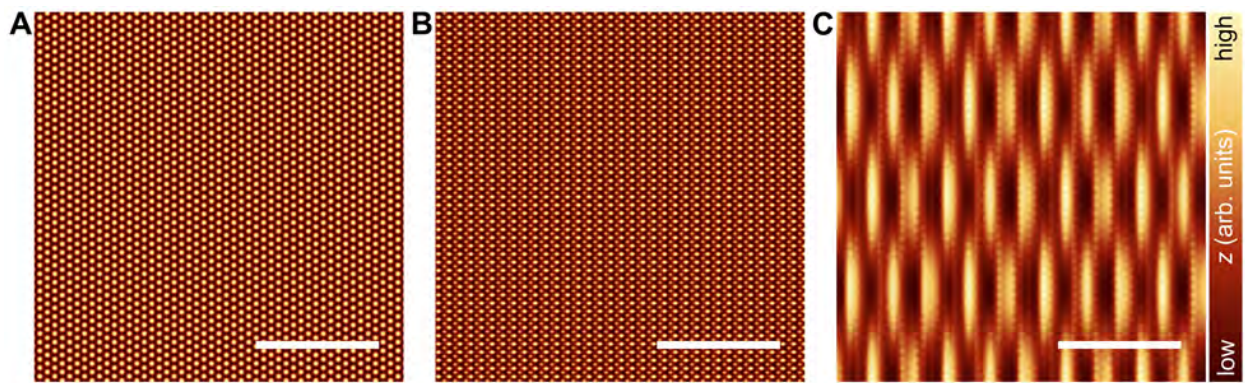

**Fig. S2. Simulation of the moiré pattern.** (**A, B**) Simulated real space image of Pb and BP obtained using a combination of sinusoidal wave functions. (**C**) The Moiré pattern obtained by a convolution of images (A) and (B) followed by smoothing using a two-dimensional Gaussian filter. (Scale bar in each panel = 5 nm).

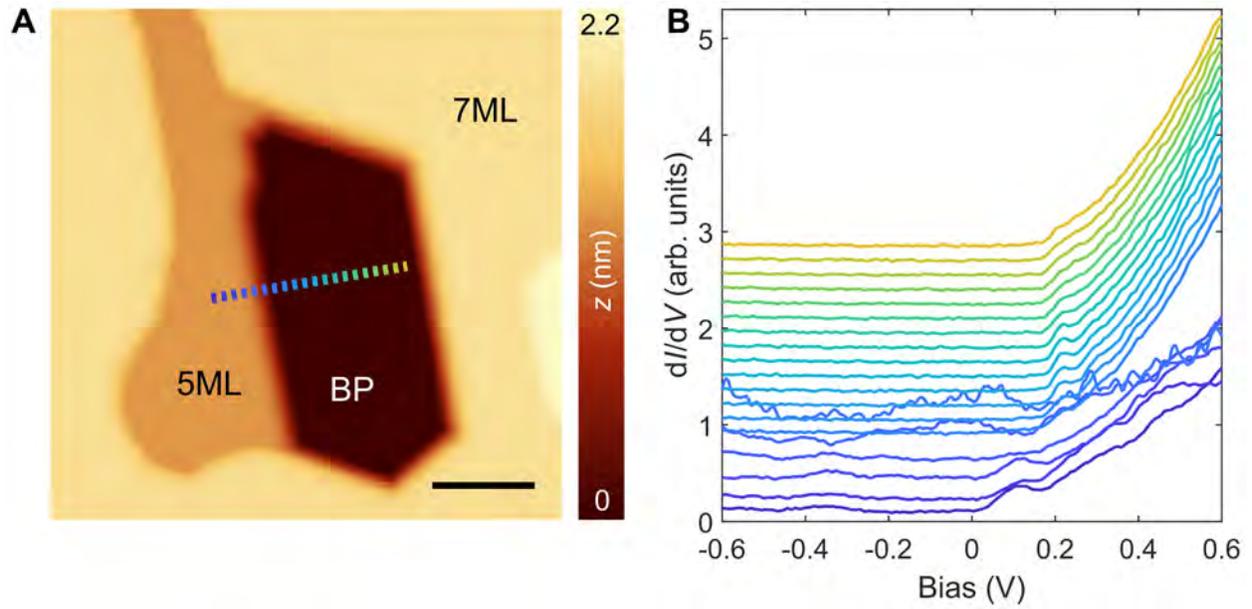

**Fig. S3. Spectroscopy across the Pb film and BP.** (**A**) Constant-current STM image of a Pb film showing various film thicknesses together with a hole where the bare BP is exposed. (**B**) d$I$/d$V$ spectra measured across the step from a 5 ML Pb film onto bare BP as shown in (A). Spectra are shifted vertically for clarity. Spectra on the 5 ML Pb film show typical spectral features corresponding to wide range spectroscopy (Fig. 3C). Spectra on BP show a bandgap around the Fermi energy with the conduction band onset at ~150 mV indicating no induced proximity effect in the substrate.

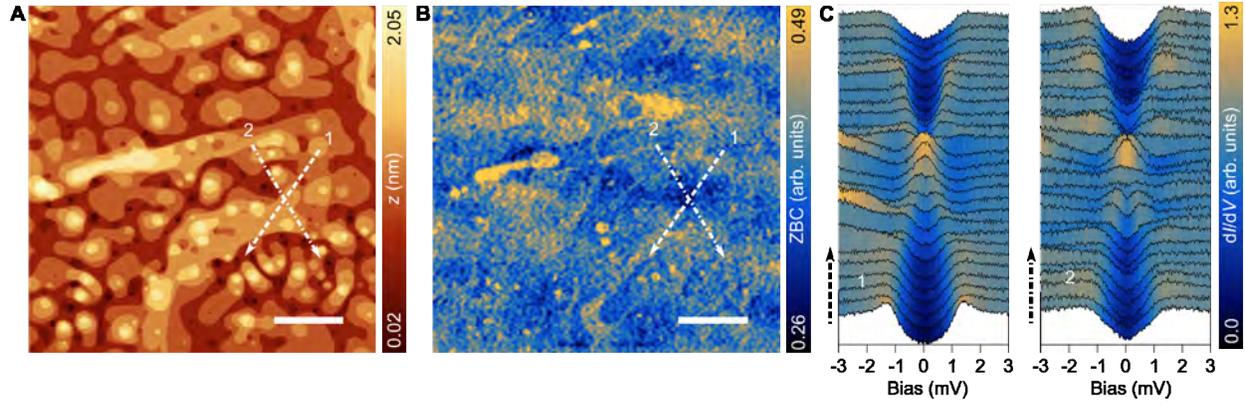

**Fig. S4. Vortex imaging of a 30ML Pb film.** (**A**) Constant-current STM image of a thick Pb film (~30 ML) showing the topography of the area used for vortex imaging. (**B**) Zero bias conductance (ZBC) map at $T$ = 30 mK and $H$ = 50 mT showing a diffused vortex lattice. Imaging parameters for (A) and (B): $V_{stab}$ = 10 mV, $I_{stab}$ = 10 pA, $V_{mod}$ = 200 µV, $\Delta z$ = - 80 pm (scale bar = 100 nm). (**C**) d$I$/d$V$ spectra measured across a vortex in two directions (1,2) along the lines through the vortex in (B), each 183 nm in length. ($V_{stab}$ = 5 mV, $I_{stab}$ = 200 pA, $V_{mod}$ = 50 µV).

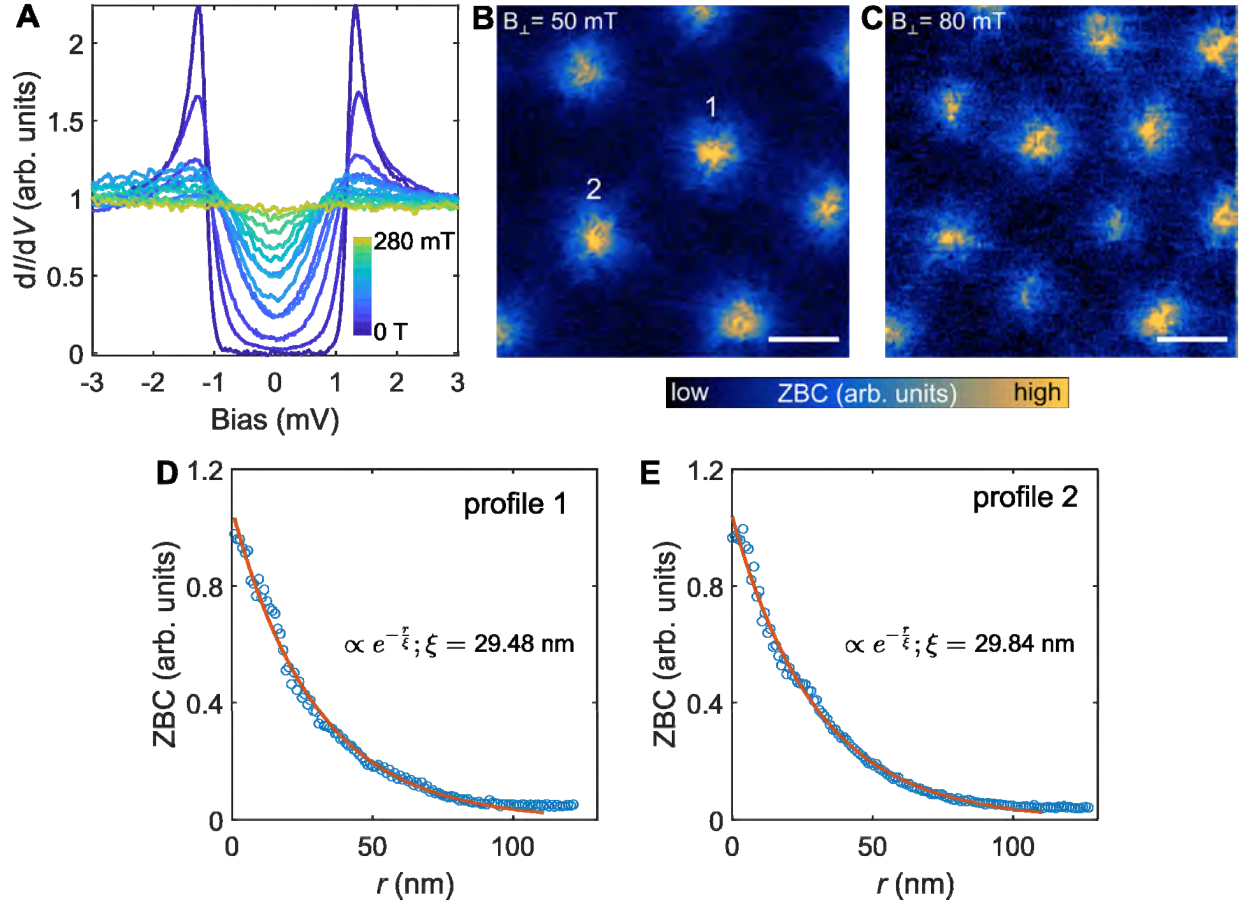

**Fig. S5. Measurement of the upper critical field and the coherence length.** (**A**) d$I$/d$V$ spectra measured as a function of out-of-plane magnetic field, measured away from vortices for as grown 7.3 ML Pb film. The superconducting gap vanishes at $H_{c2}$ = 280 mT. We estimated the coherence length using $\xi = \sqrt{\Phi_0/2\pi H_{c2}(0)}$ = 34.3 nm, where $\Phi_0$ is the quantum of magnetic flux ($V_{stab}$ = 5 mV, $I_{stab}$ = 200 pA, $V_{mod}$ = 50 μV, $T$ = 35 mK). (**B, C**) Zero bias d$I$/d$V$ map at $B_\perp$ = 50 mT and $B_\perp$ = 80 mT, showing an Abrikosov vortex lattice (panel (B) is same as Fig. 2B of the main manuscript). Imaging parameters for (B) and (C): $V_{stab}$ = 10 mV, $I_{stab}$ = 10 pA, $V_{mod}$ = 200 uV, $\Delta z$ = - 80 pm, scale bar = 100 nm. (**D, E**) Radially averaged profiles across two vortices marked in (B). From an exponential fit of these profiles, the coherence length could be extracted: $\xi \sim$ 30 nm. This is consistent with the value determined using the upper critical field ($H_{c2}$).

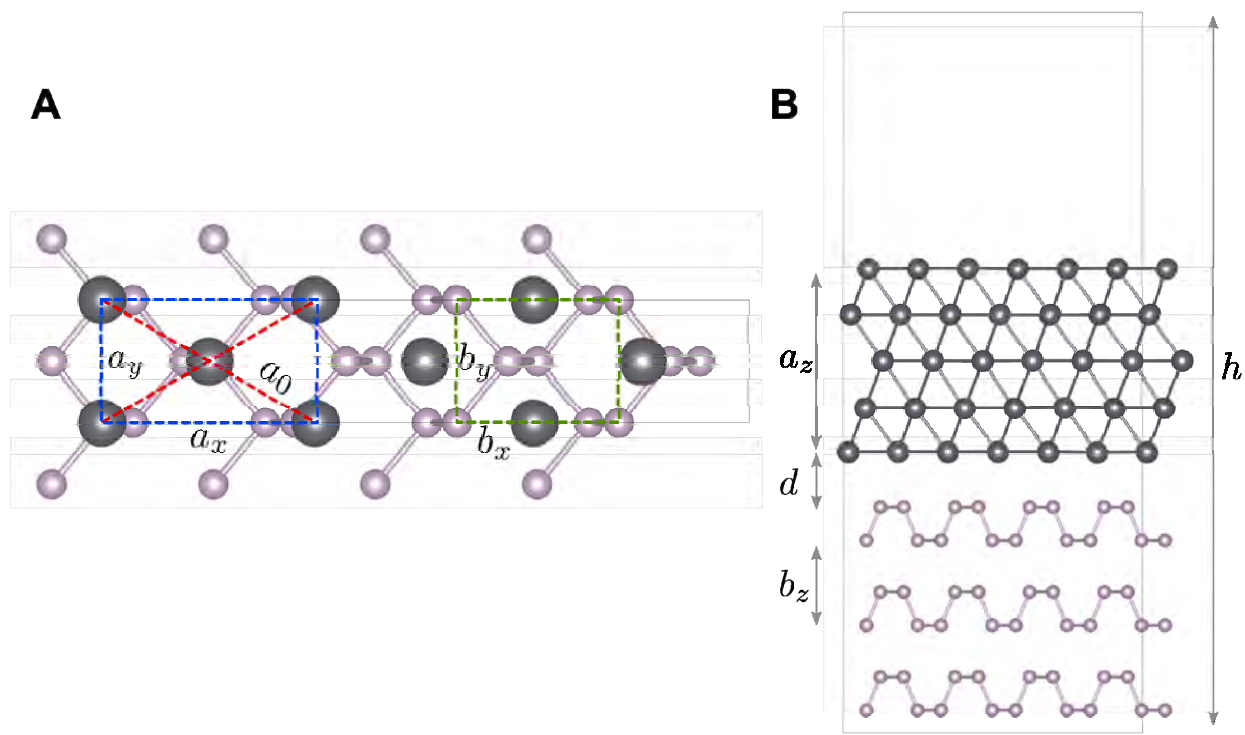

**Fig. S6. Lattice structure of the utilized supercell.** (**A**) Top view of a supercell with a monolayer Pb on a monolayer of BP. Red, blue, and green dashed lines indicate the primitive (hexagonal) Pb(111), the rectangular Pb, and the primitive BP unit cells, respectively. (**B**) Side view of a supercell will 5 ML of Pb on top of 3 ML of BP.

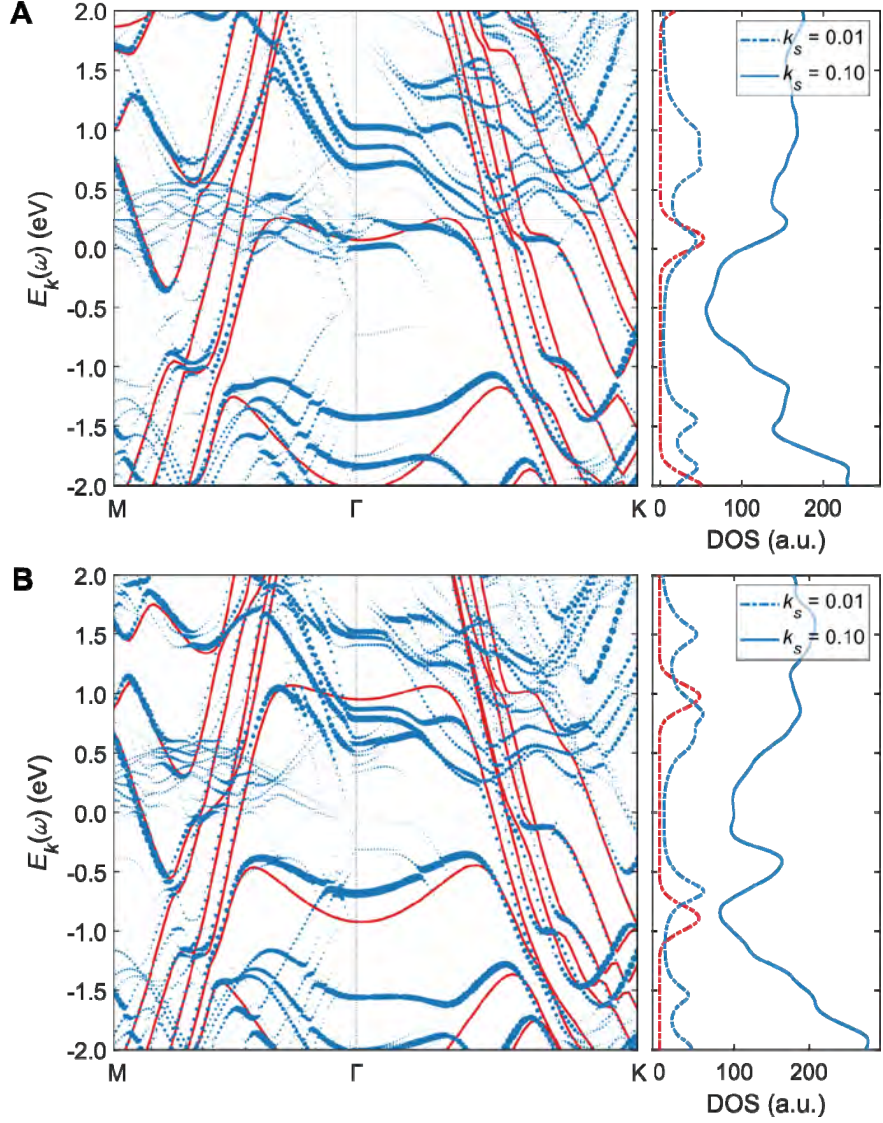

**Fig. S7. Unfolded band structures and weighted density of states a Pb film on three layers of BP.** In (A) and (B) data for 5ML and 6ML Pb on 3ML BP is shown. The band structures are plotted along high-symmetry points of the primitive Pb (111) unit cell. Solid red lines in the band structures refer to the pristine (but strained) 5ML and 6ML Pb band structure. Blue dots correspond to the unfolded spectral function as defined in Eq.(4) where the dot-size represents the intensity of A(k,w) at the corresponding momentum and energy. The side panels of (A) and (B) show the quasi-local spectral functions $\rho(\omega)$ using different smearing $\sigma_k$. Blue lines represent unfolded heterostructure data, while the red lines depict $\rho(\omega)$ for the pristine Pb structures.

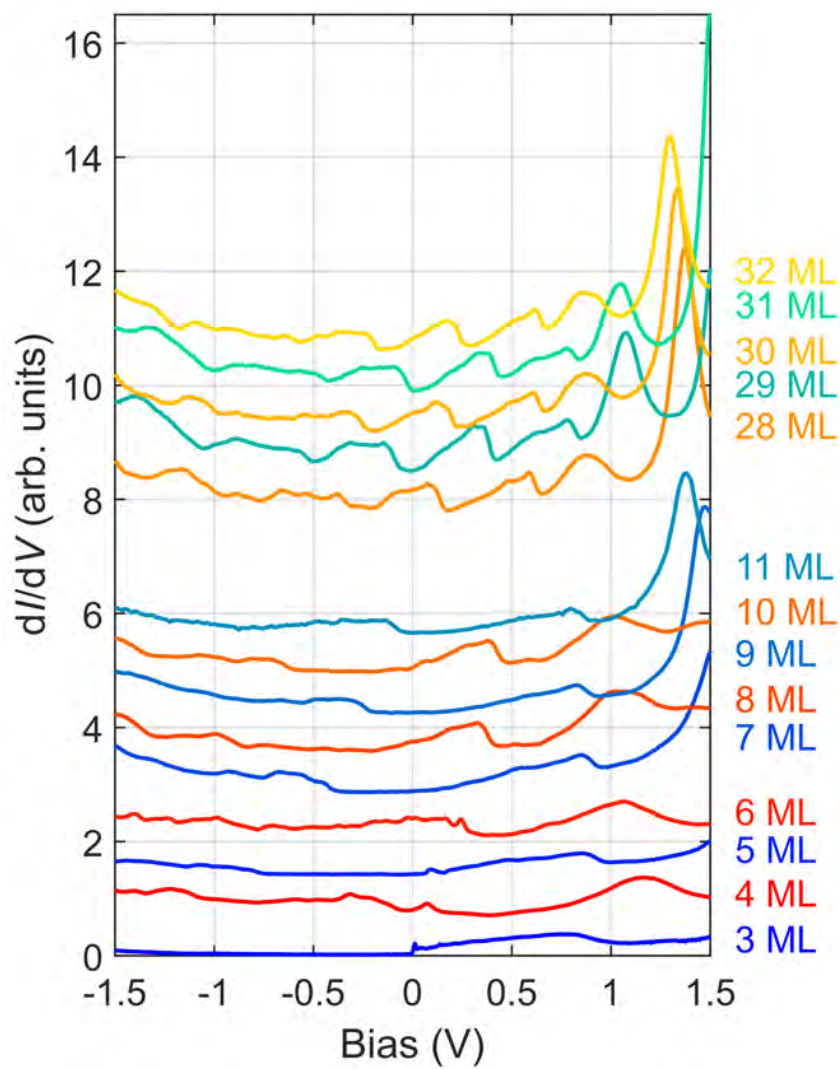

**Fig. S8. Evolution of the LDOS with varying film thickness.** Series of d$I$/d$V$ spectra measured in a wide bias range for Pb films of thicknesses indicated on the right. Even/odd films are plotted with same color shades (red/blue). Spectra are shifted vertically for clarity (stabilization parameters for 3-6 ML: $V_{stab}$ = 1.5 V, $I_{stab}$ = 500 pA, $V_{mod}$ = 5 mV; for 7-32 ML: $V_{stab}$ = 1.5 V, $I_{stab}$ = 200 pA, $V_{mod}$ = 10 mV. Measurement temperature for 3-6 ML, 28-32 ML: $T$ = 35 mK; measurement temperature for 7-11 ML: $T$ = 1.3 K).

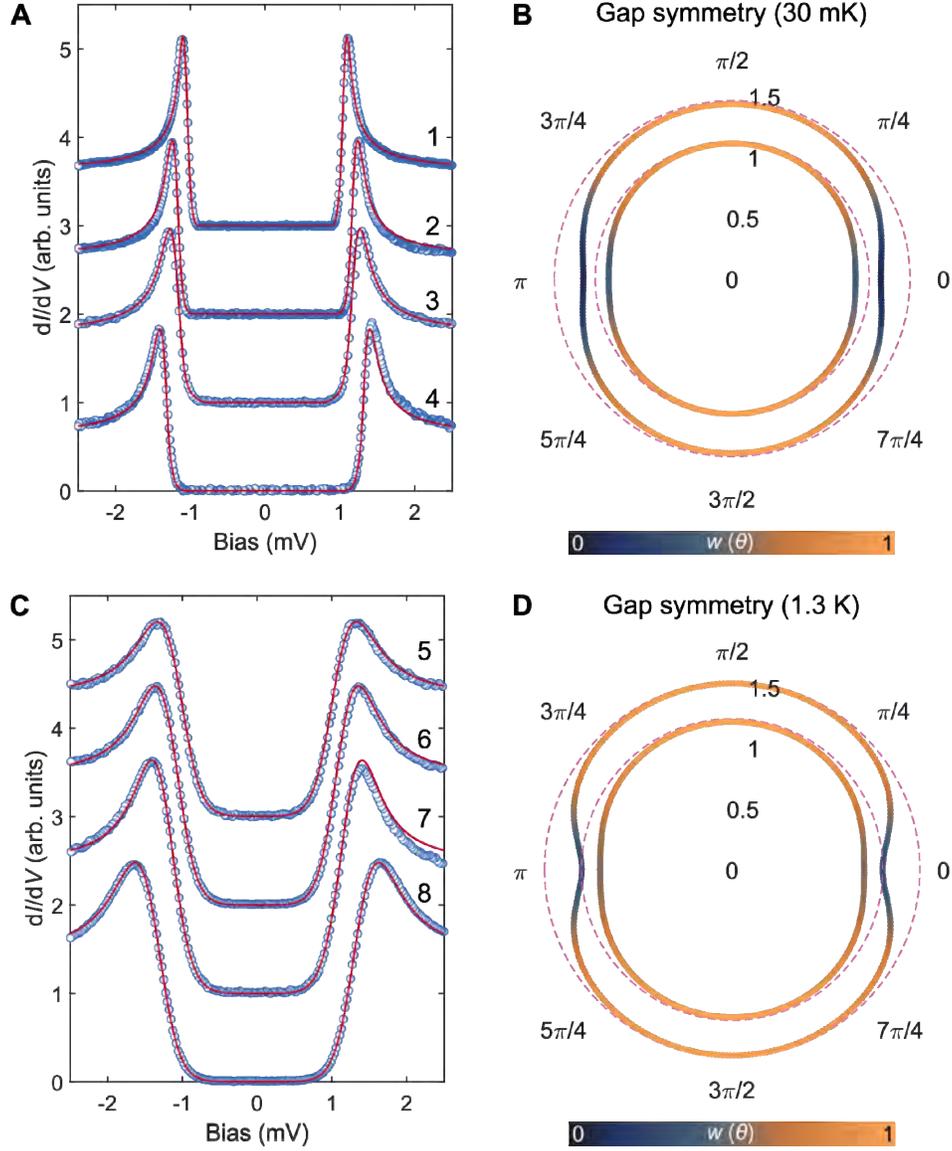

**Fig. S9. Modeling superconducting gaps using a two-band hybrid model.** (**A**,**C**) d$I$/d$V$ spectra (open blue circles) measured at $T$ = 30 mK and $T$ = 1.3 K respectively on different samples with similar layer thicknesses (6-7 ML). Spectra are shifted vertically for clarity. The solid red lines in (A) and (C) are fits to the corresponding spectra using an anisotropic gap and the weight function ($w(\theta)$) as shown in the polar plots in (B) and (D) respectively. For clarity, (B) and (D) show the gap symmetry for large (outer curve) and small (inner curve) gaps only. Stabilization parameters: $V_{stab}$ = 5 mV, $I_{stab}$ = 200 pA, $V_{mod}$ = 20-100 µV. The fitting parameters used in (A) and (C) are summarized in Table S1.

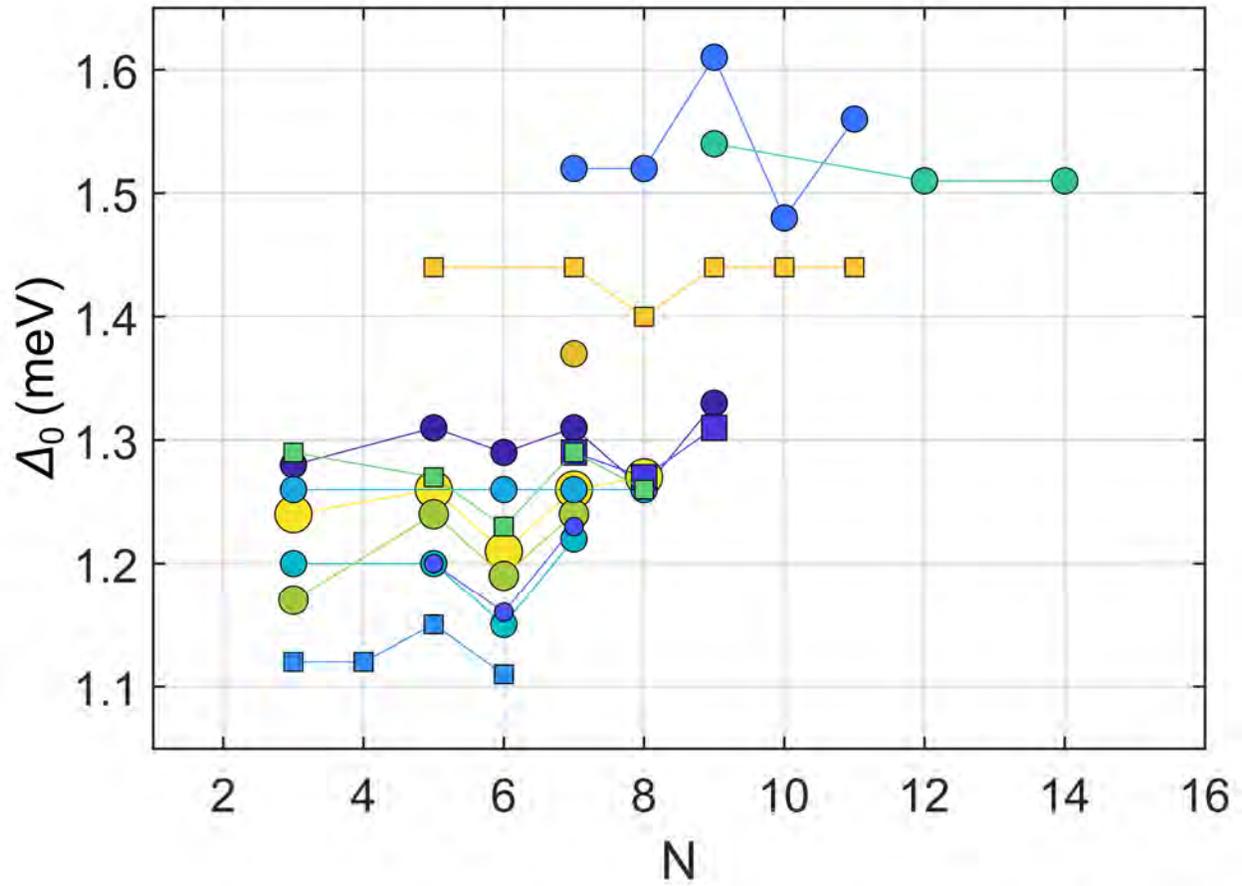

**Fig. S10. Thickness-dependence of the superconducting gap.** The size of the anisotropic gap extracted from numerical fits is plotted as a function of layer thickness for different growths. Filled squares (circles) correspond to data measured at $T$ = 30 mK ($T$ = 1.3 K). For clarity, all points within the same growth are connected with solid lines.

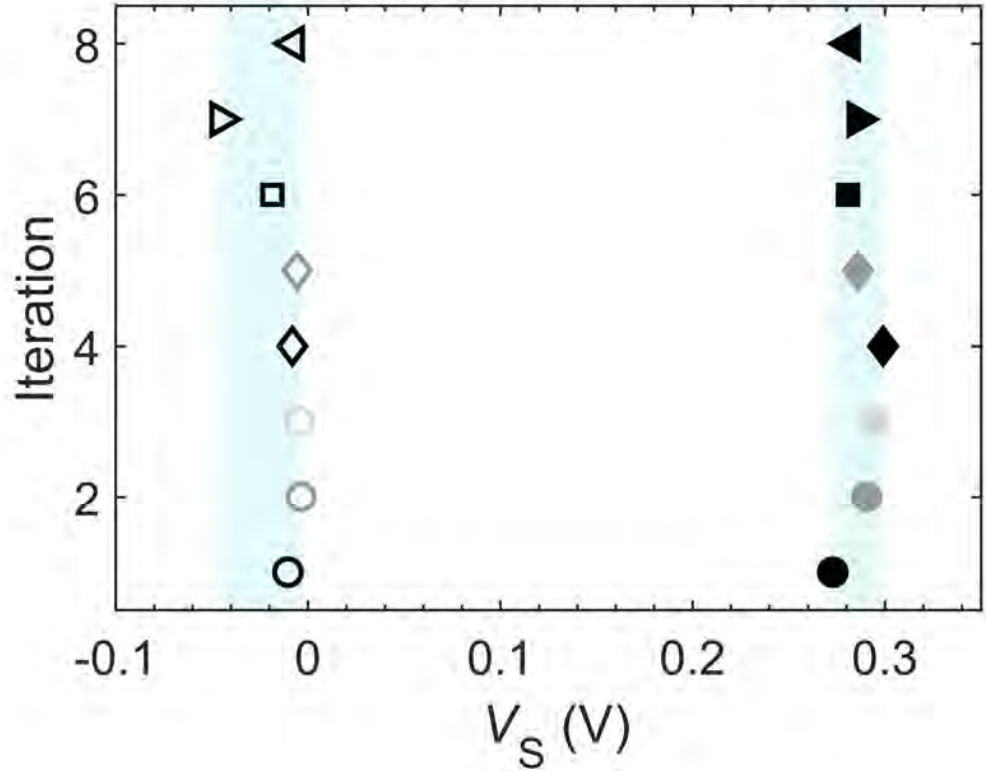

**Fig. S11. Bandgap measurement of pristine BP samples.** The plot shows the conduction band edges (filled symbols) and valence band edges (empty symbols) for different BP samples determined from d$I$/d$V$ spectra measured on the cleaved surface at $T$ = 4.4 K (*15-17*). Multiple cleaves on the same BP crystal are shown with the same symbols. The light blue color in the background represents the variation in the position of the band edges (Stabilization parameters for d$I$/d$V$ spectra: $V_{stab}$ = -200 mV, $I_{stab}$ = 20-200 pA, $V_{mod}$ = 4-10 mV).

**Table S1. Parameters used to fit the spectra in Fig. S9.**
The table summarizes different parameters used for fitting spectra in Fig. S9 as described in the supplementary note S6.

| Spectrum | Measurement T (K) | $\Delta_0$ (meV) | $\zeta$ | $\epsilon$ | $p_0$ | $\tau$ | $\chi$ | $\sigma$ |
|---|---|---|---|---|---|---|---|---|
| 1 | 30 m | 1.11 | 0.005 | 10 | 2 | 0.1 | -1 | -5 |
| 2 | 30 m | 1.29 | 0.009 | 10 | 2 | 0.2 | -0.5 | -4.5 |
| 3 | 30 m | 1.29 | 0.015 | 10 | 1.5 | 0.15 | -0.9 | -1.5 |
| 4 | 30 m | 1.44 | 0.01 | 10 | 2 | 0.15 | -0.9 | -3.8 |
| 5 | 1.3 | 1.22 | 0.035 | 10 | 1.5 | 0.1 | -0.6 | -3.1 |
| 6 | 1.3 | 1.26 | 0.017 | 10 | 1.5 | 0.1 | -0.6 | -2.3 |
| 7 | 1.3 | 1.31 | 0.002 | 10 | 1.5 | 0.1 | -0.6 | -1.4 |
| 8 | 1.3 | 1.52 | 0.025 | 10 | 1.5 | 0.1 | -1.2 | -1.5 |